\documentclass{aip-cp}

\usepackage[numbers]{natbib}
\usepackage{rotating}
\usepackage{graphicx}

\def\micro{\mu}

\def\gtorder{\mathrel{\raise.3ex\hbox{$>$}\mkern-14mu
 \lower0.6ex\hbox{$\sim$}}}
\def\ltorder{\mathrel{\raise.3ex\hbox{$<$}\mkern-14mu
 \lower0.6ex\hbox{$\sim$}}}
\def\mugegm{\mu_p G_E^p / G_M^p}

\def\gep{G_E^p}
\def\gmp{G_M^p}

\begin{document}

% Title portion
\title{Super-Rosenbluth Measurements with Electrons and Positrons}

\author[uva]{Mikhail Yurov}
%\eaddress[url]{http://www.aip.org}
\author[anl]{John Arrington\corref{cor1}}
%\eaddress{johna@anl.gov}

\affil[uva]{University of Virginia, Charlottesville, VA}
\affil[anl]{Physics Division, Argonne National Laboratory, Argonne, IL}
\corresp[cor1]{johna@anl.gov}

\maketitle

\begin{abstract}
Precise measurements of the proton form factor ratio $\mugegm$ from Rosenbluth separation
measurements can be combined with Polarization based extractions to provide significant
constraints on two-photon exchange contributions to the elastic e-p cross section.
We present an overview of JLab experiment E05-017, the high-precision 'Super-Rosenbluth'
measurements of the proton form factor taken in Hall C of Jefferson Lab. We then examine
what precision could be obtained for Super-Rosenbluth measurements using a low-intensity
positron beam at Jefferson Lab.
\end{abstract}

% Head 1
\section{INTRODUCTION}

Historically, the proton electromagnetic form factors have been measured in unpolarized
elastic e-p scattering, using the Rosenbluth separation technique. This technique has limited
sensitivity to the charge form factor, $\gep$, at large $Q^2$, and so recoil polarization
techniques have been used to extend measurements of $\gep$ to large $Q^2$ at Jefferson
Lab~\cite{punjabi05, puckett12, puckett17}. However, the polarization measurements yielded
a significant decrease in $\mugegm$ with increasing $Q^2$, while the Rosenbluth results
indicated a roughly constant ratio~\cite{arrington07a, perdrisat07, arrington11b, punjabi15}.
Initial attempts to understand the discrepancy~\cite{arrington03a, arrington04a} suggested
that it was a systematic difference, and could not be explained by an error in one or two
experiments, and a high-precision 'Super-Rosenbluth' separation~\cite{qattan05}, using
proton rather than electron detection, provided Rosenbluth extractions of the form factor
ratio with precision comparable to the polarization measurements, confirming the discrepancy.
Figure 1 shows a comparison of polarizations measurements in blue and Rosenbluth measurements
in red.

\begin{figure}[h]
  \centerline{\includegraphics[width=220pt]{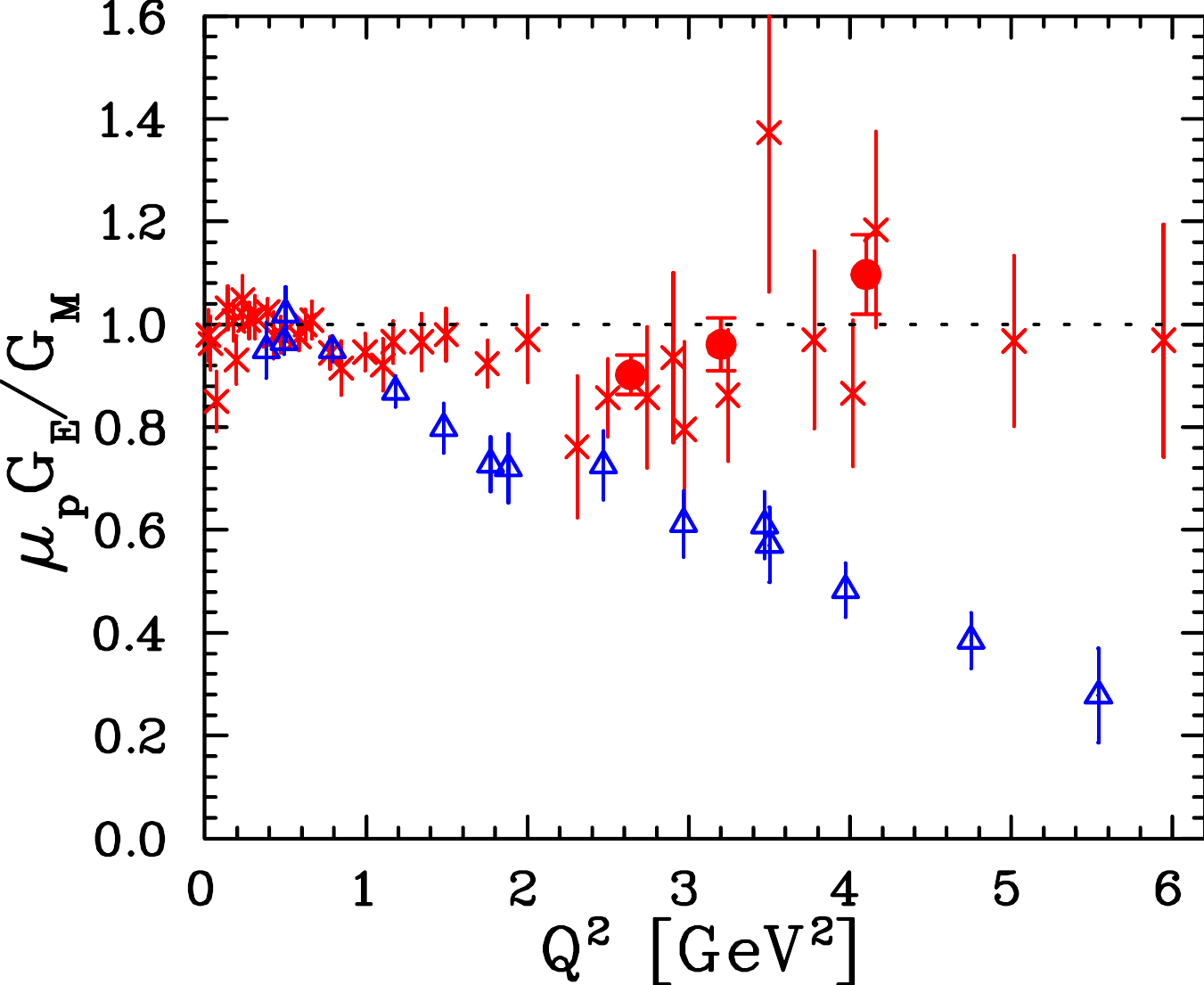}}
  \caption{Comparison of Rosenbluth and polarization extractions of $\mugegm$. The blue triangles
are from polarization transfer extractions~\cite{jones00, gayou02}, the red crosses are from a
global analysis of Rosenbluth measurements~\cite{arrington04a}, and the solid red circles are
from the E01-001 `Super-Rosenbluth' experiment~\cite{qattan05} discussed below. Figure adapted
from Ref.~\cite{qattan05}}
\end{figure}

The discrepancy is now generally attributed to the contribution of two-photon
exchange (TPE) corrections~\cite{guichon03, blunden03, afanasev05a, arrington07c, arrington11b,
afanasev17}, which have minimal impact on polarization measurements but can have a significant
impact when extracting $\gep$ from the small, angle-dependent contribution to the unpolarized
cross section. A combined analysis of existing electron-proton and positron-proton scattering
comparisons~\cite{arrington04b}, where the TPE contribution is expected to change sign with the
sign of the lepton, provided some evidence for an angle-dependent correction.
However, the indication for TPE contributions was only
observed at the 3-sigma level, and only when combining all world's data for $Q^2<2$~GeV$^2$ - 
below the region where a clear discrepancy in the form factor extractions was observed.
New measurements of electron-proton and positron-protons scattering~\cite{moteabbed13,
adikaram15,  rachek15, rimal17, henderson17} provide indications of TPE contributions
for $Q^2$ up to 2~GeV$^2$.
The data provide evidence of TPE contributions and are consistent with calculations of
the TPE contributions in hadronic models~\cite{blunden05, borisyuk06} but the precision
of the data and the limited $Q^2$ range provide limited ability to validate these 
calculations~\cite{afanasev17}, and further measurements at higher $Q^2$ are needed.

If the discrepancy between Rosenbluth and Polarization measurements is explained entirely by
TPE contributions and these contributions are nearly linear in $\varepsilon$~\cite{tvaskis06},
the combination of cross section and polarization measurements can be used to extract the form
factors without being dominated by uncertainty in the TPE corrections~\cite{arrington07c, ye17}.
However, these assumptions must be tested, and for other high-precision measurements, in 
particular at low $Q^2$ values~\cite{afanasev05b, arrington07b, tjon09, zhou10, arrington13},
we must rely on calculated corrections and so it is important
to test calculations of the TPE contributions.  

An examination of world's unpolarized cross section data was consistent with the linear dependence
on $\varepsilon$ that is expected in the single photon (Born) approximation. It also provided
limits on deviations from this linear behavior coming from contributions beyond the Born
approximation, such as TPE contributions. While the analysis set significant limits on 
non-linear contributions~\cite{tvaskis06}, these were not precise enough to rule out many
of the calculations of TPE corrections, which typically predict relatively modest deviations
from linearity. The best limits on non-linear contributions from TPE come from the initial
Super-Rosenbluth experiment~\cite{qattan05}. These data also provided measurements of
$\mugegm$ from 2.6-4.1~GeV$^2$ with precision comparable to polarization measurements, allowing
for the most precise extraction of the TPE contribution to the cross section from comparisons of
Rosenbluth and Polarization data~\cite{arrington05, qattan11, qattan15}. With the inclusion of
measurements of the polarization ratio as a function of $\varepsilon$~\cite{meziane11}, it is
possible to make similar, but still model-dependent, extractions of the full TPE
amplitudes~\cite{borisyuk11, guttmann11}.

\section{THE SUPER-ROSENBLUTH TECHNIQUE}

\begin{figure}[h]
  \includegraphics[width=150pt]{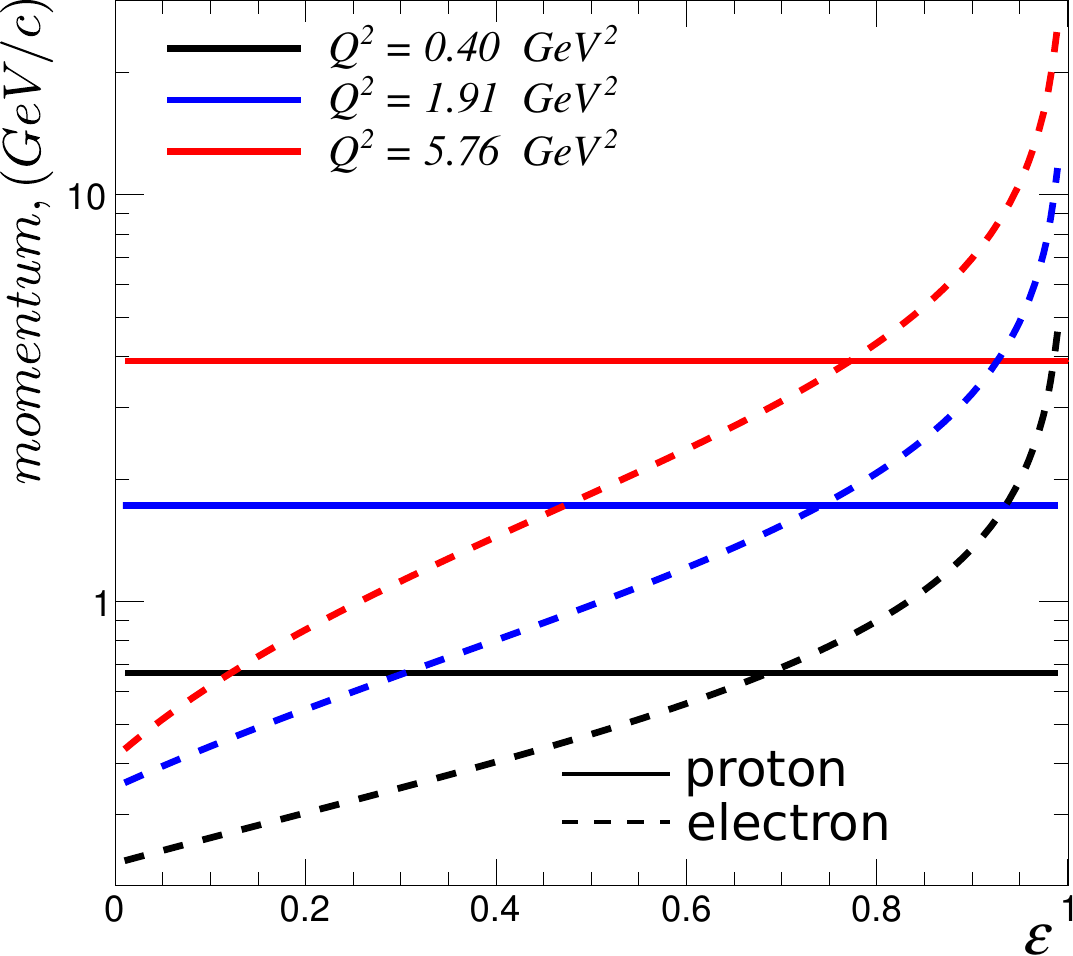}
  \includegraphics[width=150pt]{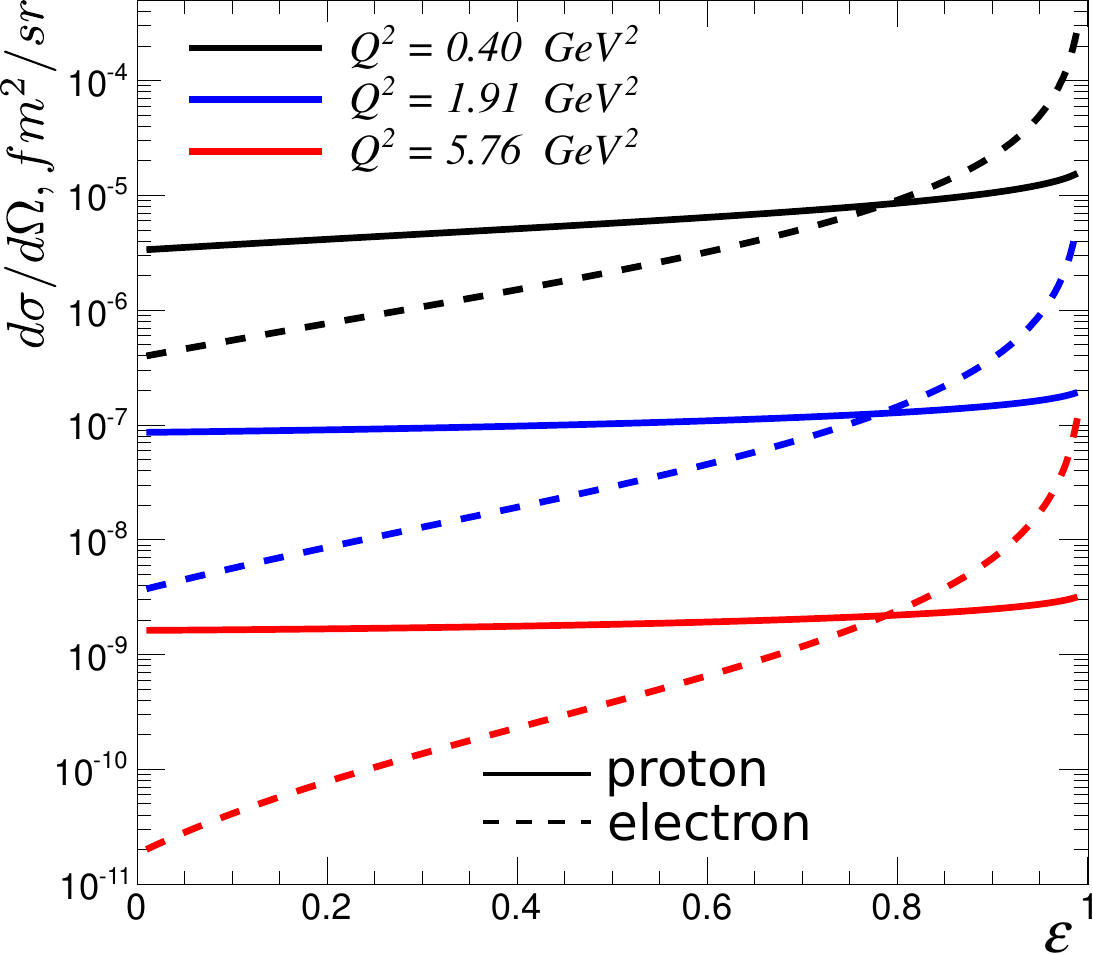}
  \includegraphics[width=150pt]{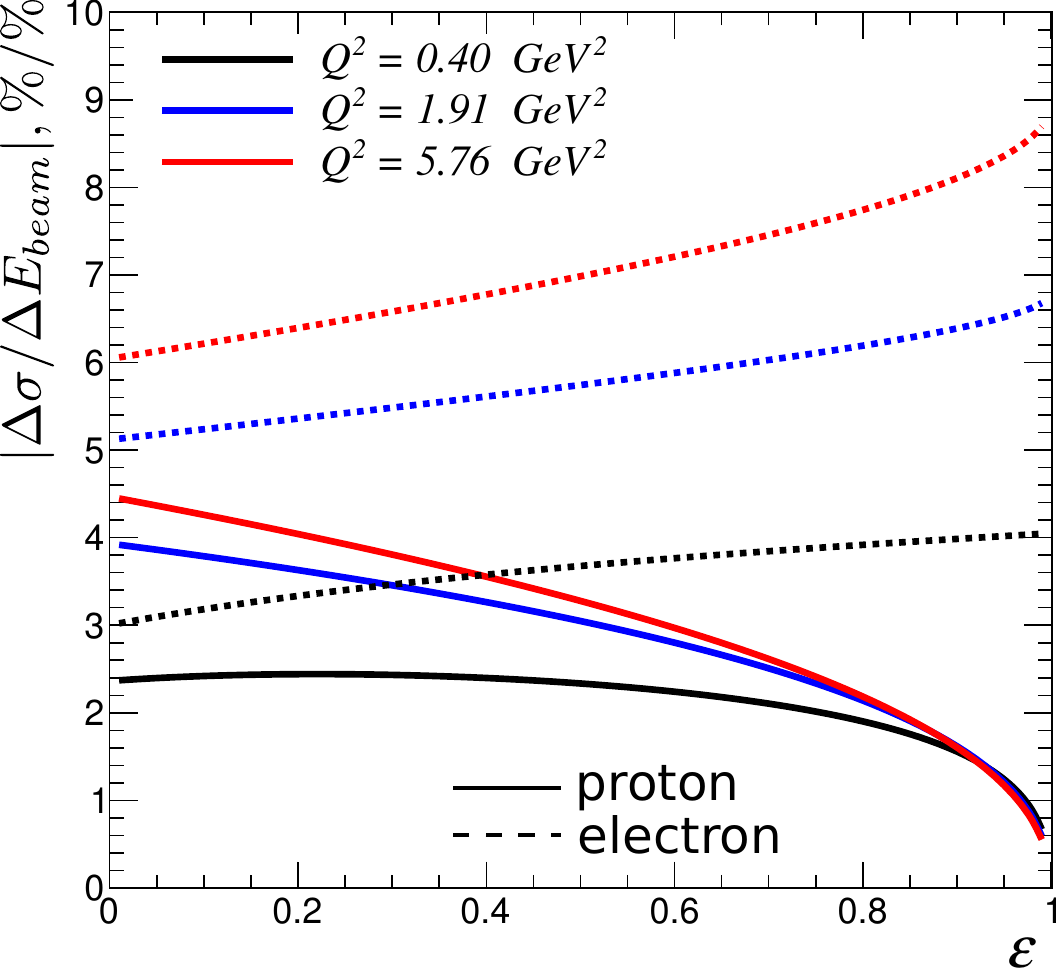}
  \caption{Detected particle momentum (left panel), cross section (middle panel), and sensitivity
to beam energy (right panel) for $Q^2$ values of 0.4, 1.91, and 5.76~GeV$^2$ for proton detection
(solid lines) and electron detection (dashed lines). At fixed $Q^2$, the proton momentum is fixed,
while the electron momentum varies over an order of magnitude to cover a $\varepsilon$ range from
0.2-0.8. The cross section for proton detection is nearly constant, while the electron cross section
varies by 1-2 order of magnitude. In addition, the proton cross section at $\varepsilon=0.2$ is a 
factor of 5-20 larger than for electron detection. The sensitivity of the cross section to a 
possible shift in beam energy is also smaller for proton detection.}
\end{figure}

JLab experiment E01-001~\cite{qattan05} made high-precision measurements of $\mugegm$ using
the Rosenbluth technique to provide confirmation of the Rosenbluth-Polarization discrepancy,
and to provide better sensitivity to deviations from the linear behavior required in the Born
approximation. The improved precision was achieved by minimizing the relative uncertainty between
points taken at different values of $\varepsilon$ but at the same $Q^2$. In a conventional
Rosenbluth separation experiment, data at fixed $Q^2$ but different $\varepsilon$ values are 
taken by making measurements at different beam energy and varying the electron scattering angle
to maintain a fixed $Q^2$ value. This means that the beam energy, detected electron energy,
and cross sections change significantly when covering a large range in $\varepsilon$. As such,
any corrections that depend on event rate or particle momentum must be corrected for precisely
or they will introduce a false $\varepsilon$ dependence. In addition, the cross section drops
rapidly as the scattering angle increases (to achieve small $\varepsilon$ values), limiting the
statistics and often leading to use of higher beam currents for the low cross section
kinematics, potentially introducing additional corrections that depend on the beam current.

JLab experiment E01-001 attempted to minimize these corrections by detecting protons rather than
electrons and by applying a tight solid angle cut to limit the measurement to the very high
acceptance region of the High Resolution Spectrometer (HRS)in Hall A. At fixed $Q^2$, protons
have fixed momentum and nearly
identical distributions in the detector, minimizing variation in efficiency or detector
response, while limiting events to a small solid angle limits potential changes in acceptance
for an extended cryogenic target. In addition, the cross section for proton detection has very
little dependence on the proton angle, meaning that data can be taken at fixed beam current to
minimize current-dependent corrections in the target, the rates in the spectrometer stay nearly
constant over the full $\varepsilon$ range, and the cross section for small $\varepsilon$ values
are significantly higher than for electron detection. Finally, the $\varepsilon$ dependence of
radiative corrections and the sensitivity of the cross section to small changes in beam energy or
scattering angle are generally smaller for proton detection. Figure 2 compares detected particle
momentum, cross section, and sensitivity to beam energy for electron and proton detection for
a selection of $Q^2$ values from 0.4 to 5.8~GeV$^2$.

\begin{figure}[h]
  \centerline{\includegraphics[width=250pt]{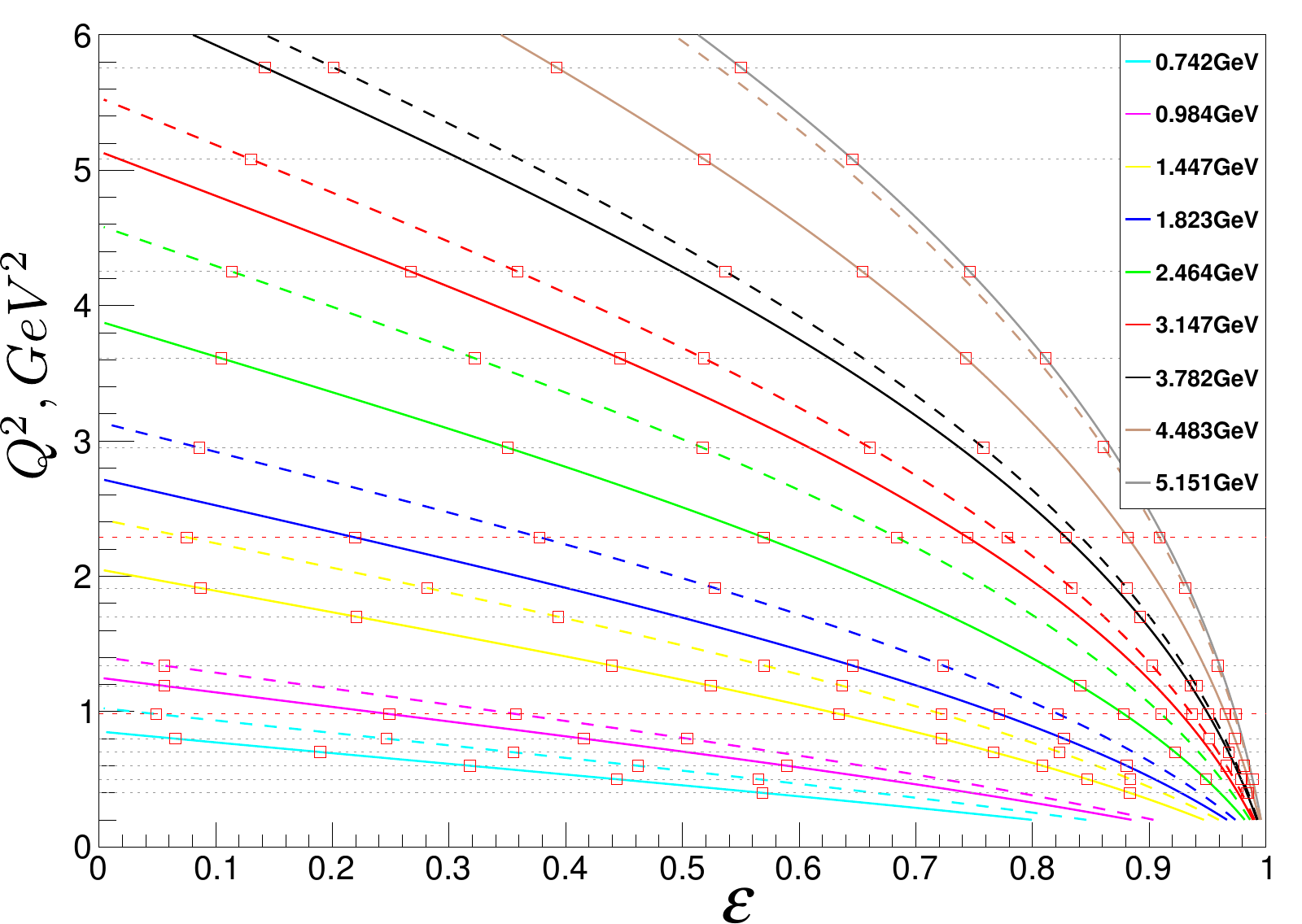}}
  \caption{Kinematics of the E05-017 experiment. The solid and dashed lines indicate beam energies
used in the measurement, with the solid lines corresponding to the energies given in the label.
The short-dashed horizontal lines indicate fixed $Q^2$ values where LT separations were performed
and the red boxes indicate the specific kinematics where elastic cross section measurements were
made.}
\end{figure}

Experiment E05-017 used the High Momentum Spectrometer (HMS) in Hall C with the goal of expanding
the $Q^2$ range of precise
Super-Rosenbluth measurements (from 0.4-5.8~GeV$^2$), while also increasing the $\varepsilon$ 
range and density of points to better constrain non-linear TPE contributions. Figure 3 shows
the beam energies and kinematic points measured in the E05-017 experiment. Analysis of the data
is nearly complete.

\subsection{Super-Rosenbluth measurements with positrons}

The Super-Rosenbluth technique is extremely well suited for making comparisons of electron-proton
and positron-protons scattering at Jefferson Lab. Because we are detecting the struck protons,
we do not have to change the polarity of the spectrometer used to detect the protons, meaning that
we do not have to worry about the backgrounds changing when switching from 
electron to positron detection. In addition, because the technique focuses on minimizing
the relative uncertainty between fixed $Q^2$ measurements at different $\varepsilon$ values,
precise extractions of $\mugegm$ can be obtained independently for positron and electron
scattering, and potential differences between the measurements should cancel in the extraction
of the ratio. For example, while positron beam measurements will have much lower event rates, due
to use of a lower intensity positron beam, this rate is nearly constant as a function of
$\varepsilon$ and so rate-dependent effects will cancel in the extraction of $\mugegm$ before
comparing electron and
positron results, as any normalization uncertainty impacts both $\gep$ and $\gmp$ identically.
Thus, the issue of precisely normalizing the luminosity between electron and positron beams
is of less significance than for other measurements. Similarly, the experiment is not sensitive
to changes in detector performance between electron and positron runs, as long as the conditions
are stable within the individual measurements. Finally, the significantly increased cross
section for low-$\varepsilon$ measurements means that one can still make measurements at
relatively large $Q^2$ values even with a significantly reduced beam current. 

We take the run times and statistics achieved in the E05-017 measurement and use this to 
make projections for the kinematic coverage achievable in a very modest (18 day) run with
a beam current of 1~$\micro$A. Because we detect protons, the only change with a positron
beam is the reduction of beam current, from 30-80~$\mu$A for E05-017 to an assumed 1~$\mu$A.
The E05-017 measurement is generally systematics limited, so we allow for statistical uncertainties
of up to 1\% at high $Q^2$, giving only a slight increase in the total uncertainty. Under
these conditions, with the same 4~cm cryotarget as used in E05-017, a positron measurement with 
1~$\mu$A unpolarized beam could cover a $Q^2$ range from 0.4-4.2~GeV$^2$ with better than 1\%
statistical uncertainty, yielding precision comparable to previous Super-Rosenbluth experiments.
With an increase to 5~$\mu$A positron current or with use of a 10~cm hydrogen target, this could
be extended to $>$5~GeV$^2$. These estimates assume that we measure at roughly half of the $Q^2$
points shown in Figure 3, and have fewer $\varepsilon$ points for some of the larger $Q^2$ values.

\begin{figure}[h]
  {\includegraphics[width=200pt]{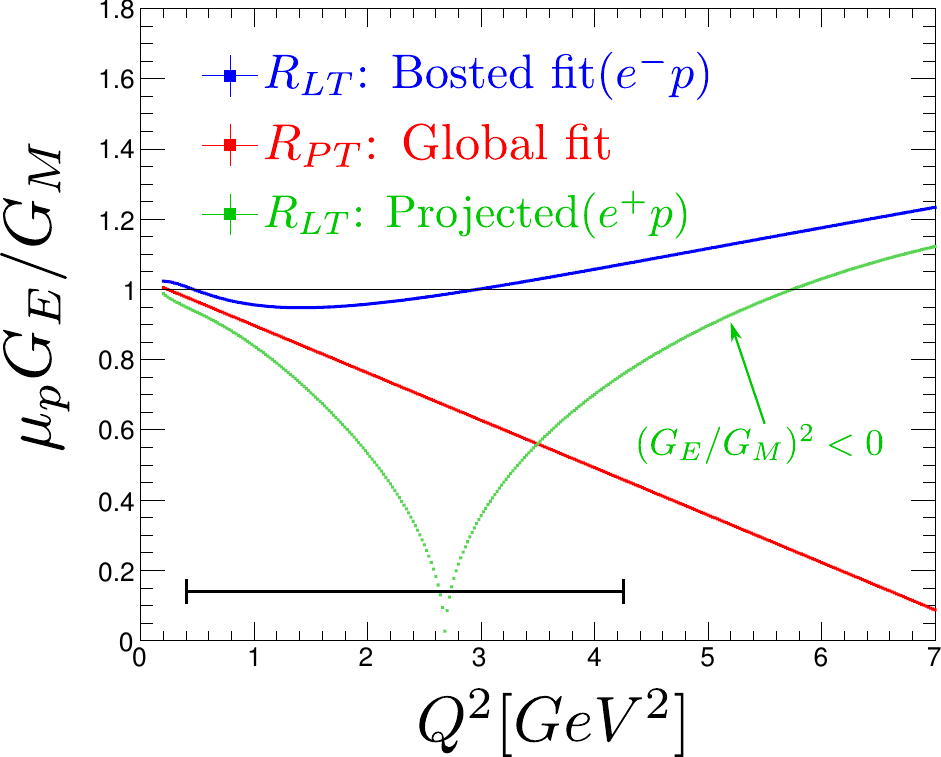}}
  {\includegraphics[width=200pt]{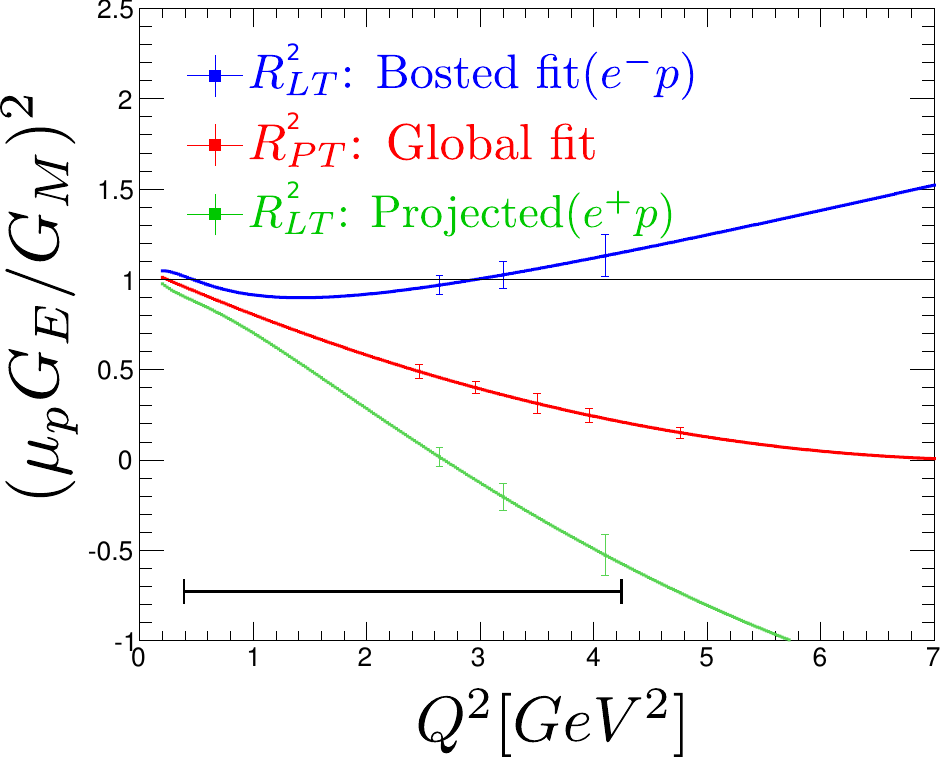}}
  \caption{The left figure shows parameterization of global Rosenbluth extractions of
$\mugegm$ (top blue line),
recoil polarization measurements (middle red line), and the expected ratio for positron measurements
(bottom green line). Note that the ratio goes to zero at $Q^2 \approx 2.6$~GeV$^2$, and becomes 
imaginary at higher $Q^2$, as $(\mugegm)^2$ goes negative. The right figure shows $(\mugegm)^2$,
for which the value is always real. The uncertainties shown on the PT and LT lines correspond
to existing data in the 2-5~GeV$^2$ region, from the Super-Rosenbluth experiments
E01-001~\cite{qattan05} for LT. The point on the projected e+p curve shows the uncertainties
based on the precision achieved in the E01-001 experiment. The black band at the bottom indicates
the range over which precision positron measurements can be made.}
\end{figure}

Figure 4 shows parameterizations of $\mugegm$ from polarization measurements and from
electron and positron Rosenbluth separations. The right figure includes the uncertainties from
the E01-001 Super-Rosenbluth measurement, with projected uncertainties for positrons assuming
comparable precision. Our estimates based on E05-017 would allow for several more $Q^2$ values
than shown in Figure 4, from 0.4-4.2~GeV$^2$, but with slightly larger uncertainties. 

Comparisons of electron and positron LT separations provide several improvements on existing 
TPE constraints. Under the assumption that TPE fully explains the Rosenbluth-Polarization difference,
the difference between the electron Rosenbluth and polarization data can be used to constrain 
the linear contribution from TPE corrections. Comparison of positron and electron scattering yields
twice the difference, improving our
sensitivity to TPE contributions. More importantly, this comparison is only sensitive to
TPE contributions, providing direct evidence that the discrepancy is caused by TPE rather than
making this assumption as a starting point for the comparison. In fact, at
$Q^2=4$~GeV$^2$, the contribution to the $\varepsilon$ dependence from $\gep$ is estimated to be
roughly one sixth of the contribution coming from TPE, meaning that measurements with positrons 
will yield an $\varepsilon$ dependence of a similar magnitude but of the opposite slope,
as can be seen in the high-$Q^2$ region of Figure 4.

In addition to comparing electron and positrion extractions of $(\mugegm)^2$ to constrain the linear
part of the TPE corrections, we can also compare the detailed $\varepsilon$ dependence of electron
and positron Rosenbluth separations to achieve improved sensitivity to non-linear contributions from
TPE~\cite{tvaskis06}.

\subsection{Conclusions}

In summary, comparisons of precise Rosenbluth separation measurements for electrons and positrons
can provide direct evidence of TPE contributions, and an improved extraction of the size of
the TPE corrections, particular at large $Q^2$ where there is a clear discrepancy between electron
Rosenbluth and polarization extractions of $\mugegm$. In addition, these measurements
would allow for improved constraints on non-linear contributions from TPE. 

Using the Super-Rosenbluth technique, where only the struck proton is measured, such precise
comparisons can be made between electrons and positrons, with no contribution to the uncertainty
coming from differences in electron and positron beam currents. Projections based on the data
from E05-017 suggest that positron LT separations can be performed for 8-9 $Q^2$ from 0.4 to
4-5~GeV$^2$, with comparable precision to the electron measurements which covered 16 $Q^2$
values from 0.4-5.8~GeV$^2$.

\section{ACKNOWLEDGMENTS}
This work was supported by the U.S. Department of Energy Office of Science, Office of Nuclear
Physics, under contracts DE-AC02-06CH11357 and DE-FE02-96ER40950.

% References
\bibliographystyle{aipnum-cp}%
\bibliography{jpos_yurov}%

\end{document}